\documentclass[12pt,a4paper]{article}

\usepackage{pstricks}
\usepackage{pst-slpe}
\usepackage{pst-blur}

\usepackage{amsfonts}
\usepackage[intlimits]{amsmath}
\usepackage{pdfcolmk}
\usepackage{mathrsfs}
\usepackage{booktabs}
\usepackage{fancyhdr}
\usepackage{epsfig}
\usepackage{verbatim}
\usepackage{color}

\usepackage{lscape}

\newcommand{\E}{{\cal E}}

\newcommand{\beq}{\begin{equation}}
\newcommand{\eeq}{\end{equation}}
\newcommand\beqa{\begin{eqnarray}}
\newcommand\eeqa{\end{eqnarray}}
\newcommand\bea{\begin{array}}
\newcommand\eea{\end{array}}
\newcommand{\nn}{\nonumber}
\newcommand{\neqa}{\nonumber\end{eqnarray}}
\newcommand{\la}{\label}
\newcommand{\J}{{\cal J}}

\renewcommand{\O}{{\cal O}}

\newcommand{\eq}[1]{(\ref{#1})}

\newcommand{\h}{\hat}
\renewcommand{\t}{\tilde}

\def\({\left(}
\def\){\right)}
\def\[{\left[}
\def\]{\right]}

\def\<{\langle}
\def\>{\rangle}

\def\d{\partial}

\def\sl{\sqrt{\lambda}}
\def\sG{/\hspace{-0.3cm}G}

\begin{document}

\begin{flushright}
LPTENS-07/15\\
hep-th/0703266
\end{flushright}
\vspace{1cm}

\begin{center}
{\Large\bf Constructing the AdS/CFT dressing factor}\\
\vspace{1cm}
Nikolay Gromov\footnote{nikgromov@gmail.com}$^{a,b}$,\hspace{0.3cm}
Pedro Vieira\footnote{pedrogvieira@gmail.com}$^{a,c}$\\
\vspace{1cm}
{\it\footnotesize $^a$ Laboratoire de Physique Th\'eorique
de l'Ecole Normale Sup\'erieure\\ et l'Universit\'e Paris-VI,
Paris, 75231, France\\ \vspace{.2cm}
$^b$ St.Petersburg INP, Gatchina, 188 300, St.Petersburg, Russia
\\ \vspace{.2cm}
$^c$
 Departamento de F\'\i sica e Centro de F\'\i sica do Porto
Faculdade de Ci\^encias da Universidade do Porto\\
Rua do Campo Alegre, 687, \,4169-007 Porto, Portugal }
\end{center}
\vspace{1cm}
\begin{abstract}
We prove the universality of the Hernandez-Lopez phase by deriving it from first principles. We find a very simple integral representation for the phase and discuss its possible origin from a nested Bethe ansatz structure. Hopefully, the same kind of derivation could be used to constrain higher orders of the full quantum dressing factor.
\end{abstract}

\newpage

\section{Introduction}

Bethe ansatz equations, first written by Bethe in the study of one dimensional metals \cite{Bethe:1931hc}, seem to be a key ingredient in the AdS/CFT duality \cite{Maldacena:1997reGubser:1998bcGubser:1998bc} between $\mathcal{N}=4$ SYM and type IIB superstring theory on $AdS_5\times S^5$.

The $\mathcal{N}=4$ SYM dilatation operator in the planar limit can be perturbatively computed in powers of the t'Hooft coupling $\lambda$. In the seminal work of Minahan and Zarembo \cite{Minahan:2002ve} it was shown that the $1$-loop dilatation operator acts on the six real scalars of the theory exactly like an integrable $SO(6)$ spin chain Hamiltonian.
Restricting ourselves to two complex scalars we obtain the same Hamiltonian considered by Bethe for the Heisemberg spin chain. The eigenstates are $K$ magnon states with momentum and energy parameterized by the roots $u_i$ which satisfy Bethe equations
\beq
\(\frac{u_i+i/2}{u_i-i/2}\)^L=\prod_{j\neq i}^K\frac{u_i-u_j+i}{u_i-u_j-i} \,.\nn
\eeq
The full $\mathcal{N}=4$ $1$-loop dilatation operator \cite{Beisert:2003jj} is also given by an integrable Hamiltonian whose spectrum is dictated by a system of seven Bethe equations \cite{Beisert:2003yb}, corresponding to the seven nodes of the $psu(2,2|4)$ Dynkin diagram.
In \cite{Beisert:2004hm} the all loop version of the Bethe equation for the $SU(2)$ sector was conjectured to be
\beqa
\(\frac{y^+_{j}}{y^-_{j}}\)^L=\prod_{j\neq i}^K\frac{u_i-u_j+i}{u_i-u_j-i}\,,\la{su2Gauge}
\eeqa
where $y_{j}(u_{j})$ and $y_{j}^\pm(u_{j})$ are given by
\beq
y+\frac{1}{y}=\frac{4\pi}{\sqrt{\lambda}} \,u \,\,\,\,\,\, , \,\,\,\,\,\, y^\pm+\frac{1}{y^\pm}=\frac{4\pi}{\sqrt{\lambda}}\(u \pm \frac{ i}{2}\) \,. \nn
\eeq
On the other hand, for the same  sector but from the string side of the correspondence, a map between classical string solutions and Riemann surfaces was proposed \cite{KMMZ}. The resemblance between the cuts connecting the different sheets of these Riemann surfaces and the distribution of roots of the Bethe equations in the scaling limit seemed to indicate that the former could be the continuous limit of some quantum string Bethe ansatz. In \cite{Arutyunov:2004vx} these equations were proposed to be
\beqa
\(\frac{y^+_{j}}{y^-_{j}}\)^L=\prod_{j\neq i}^K\frac{u_i-u_j+i}{u_i-u_j-i} \, \sigma_{\rm AFS}^2(u_i,u_j) \la{su2String} \,,
\eeqa
where
\beq \la{sigma2}\sigma_{{\rm AFS}}(u_i,u_j)=
 \frac{1-1/(y_j^+ y_i^-)}{1-1/(y_j^-y_i^+)}  \(\frac{y_j^-
y_i^--1}{y_j^-y_i^+-1} \frac{y_j^+ y_i^+-1}{y_j^+ y_i^--1}\)^{ i(u_j-u_i)}\,.
\eeq
The striking similarity between (\ref{su2Gauge}) and (\ref{su2String}) naturally led to the proposal that both sides of the correspondence would be described by the same equation which a scalar factor $\sigma^2$ interpolating from $\sigma_{\rm AFS}^2$ for large t'Hooft coupling to $1$ for small $\lambda$.

In \cite{Beisert:2005fw} Beisert and Staudacher (BS) conjectured the all-loop Bethe equations for the full $PSU(2,2|4)$ group and  in \cite{Beisert:2005tm} these equations were brought to firmer ground. As before, one of the main tools used to guess the form of these equations was the existence of the classical algebraic curve for the full superstring on $AdS_5\times S^5$ \cite{BKSZ}.
The seven equations for the seven types of roots $u_{a,j}$ are entangled and the middle equation, the most complicated one, can be written as\footnote{We choose to write it in this very general form -- this might at first seem strange in the sense that, for a generic potential $\mathcal{V}$ depending on the position of all the roots, this equation does not seem to describe a factorized scattering process. Obviously, eventually the phase must be such that this property is satisfied.}
\beqa
\(\frac{y^+_{4,k}}{y^-_{4,k}}\)^L&=& e^{-i{\cal V}(y_{4,k})}
\prod_{j\neq k}^{K_4}
\frac{u_{k,4}-u_{j,4}+i}{u_{k,4}-u_{j,4}-i}
\,
\sigma_{\rm AFS}^2(u_{4,k},u_{4,j}) \nn
\\
&& \qquad
\times
\prod_{j=1}^{K_1}
\frac{1-1/y_{4,k}^{- } y_{1,j}}{1-1/y_{4,k}^{+ }y_{1,j}}
\prod_{j=1}^{K_3}
\frac{y_{4,k}^{- }-y_{3,j}}{y_{4,k}^{+ }-y_{3,j}}
\prod_{j=1}^{K_5}
\frac{y_{4,k}^{- }-y_{5,j}}{y_{4,k}^{+ }-y_{5,j}}
\prod_{j=1}^{K_7}
\frac{1-1/y_{4,k}^{- }y_{7,j}}{1-1/y_{4,k}^{+ }y_{7,j}}\,. \la{middle}
\eeqa
Finally, the energy of the string states (or anomalous dimension of the SYM operators) is carried by the middle root only
\beqa
\Delta=\frac{\sqrt{\lambda}}{2\pi}\sum_{i=1}^{K_4}\(\frac{i}{y_{4,j}^+}-\frac{i}{y_{4,j}^-}\) \la{Delta4}\ .
\eeqa
The potential phase $\mathcal{V}$ should be responsible for the interpolation between the YM and the string equations for small and large t'Hooft coupling $\lambda$.

Based on the $1$-loop shift analysis of some classical circular strings \cite{Frolov:2003qcFrolov:2003tuArutyunov:2003zaPark:2005jiBeisert:2005cw} Hernandez and Lopez proposed a universal form for the first $1/\sqrt{\lambda}$ correction in $\mathcal{V}$ \cite{HL} which should be able to reproduce, together with the finite size corrections to the scaling limit \cite{Schafer-Nameki:2005tn,Beisert:2005mqHernandez:2005nfBeisert:2005bvGromov:2005gp}, the $1$-loop shift around \textit{any} classical solution. This was quite a bold proposal since it relied solely on the analysis of rank one rigid circular strings. In \cite{Freyhult:2006vr} the proposed phase passed a very nontrivial test -- it was shown to reproduce the non-analytic contribution to the $1$-loop shift around a simple $SU(3)$ solution. However, at the present stage, only for the $sl(2)$ one cut solution the $1$-loop shift from the Bethe ansatz equations is completely understood \cite{Schafer-Nameki:2005tn,HL}.

Recently, using the crossing constraint \cite{Janik:2006dc} and transcendentality principles \cite{Kotikov:2002ab}, the full form of the scalar factor was conjectured from the string \cite{Beisert:2006ibBeisert:2006zy} and gauge \cite{Beisert:2006ezEden:2006rx} theory points of view.

It is therefore fair to say that the advance in the last four years was spectacular. On the other hand it is also true that there is a great deal of conjectures involved one should both check and, hopefully, proof (or disproof). In this article we establish rigourously the Hernandez-Lopez phase by fully proving its universality.

Finally we discuss a possible origin for this phase from a nested Bethe ansatz approach.

\section{Brief derivation of the Hernandez-Lopez coefficients}
One can compute the fluctuation energies, i.e. the energy level spacing in the spectrum of the string, around any classical string solution in $AdS_5\times S^5$. To do so one should first map the classical configuration to an $8$--sheet Riemann surface \cite{BKSZ} described by a set of $8$ quasi-momenta $p_i(x)$. In particular the classical energy $\E_{cl}$ of the solution is encoded in the large $x$ asymptotics of the $p_i(x)$. Then, to compute the fluctuation energies $\E_{n}^{ij}$ one adds $N_n^{ij}$ poles with residue
\beq
\alpha(x)=\frac{\sqrt{\lambda}}{4\pi}\frac{x^2}{x^2-1} \nn
\eeq
to the quasi-momenta $p_i$ and $p_j$. The choice of the pairs of sheets $ij$ corresponds to the several string polarizations and the mode number $n$, the Fourier mode of the excitation, fixes the position $x_n$ of the new poles by
\cite{GV}
\beq
p_i (x_n)-p_j (x_n)=2\pi n \,. \la{map}
\eeq
Then the excitation energies can be read from the large $x$ asymptotic of the perturbed quasi-momenta \cite{GV}
\beq
\E=\E_{cl}+\sum_{n=-\infty}^{\infty} \sum_{ij} N_{n}^{ij} \E_{n}^{ij}\, \nn
\eeq
An object of prime interest obtained from these fluctuation energies is the $1$-loop shift \cite{Frolov:2002av}
\beq
\E_0=\frac{1}{2}\sum_{n=-\infty}^{\infty} \sum_{ij} (-1)^F \E_{n}^{ij} \la{1loop}
\eeq
where $(-1)^F=\pm 1$ for bosonic and fermionic excitations respectively. To sum over all fluctuation energies it is extremely useful to employ the technique used by Schafer-Nameki in \cite{Sakura} and integrate them in the $n$ plane using the poles of the cotangent to pick the integer values of the mode numbers $n$,
\beq
\E_0=\frac{1}{4i} \int_{\mathcal{C}} \cot( \pi n)\( \sum_{ij} (-1)^F \E_{n}^{ij} \)dn \la{cot}
\eeq
where $\mathcal{C}$ encircles all the poles of $\cot$ (and only them) .
The structure of singularities of the fluctuation energies in the complex $n$ plane is intricate and we shall discuss it in more detail elsewhere \cite{GV2}. For a simple $su(2)$ circular string solution this analysis was carried out in \cite{Sakura}.
\begin{figure}[t]
    \centering
        \resizebox{120mm}{!}{\includegraphics{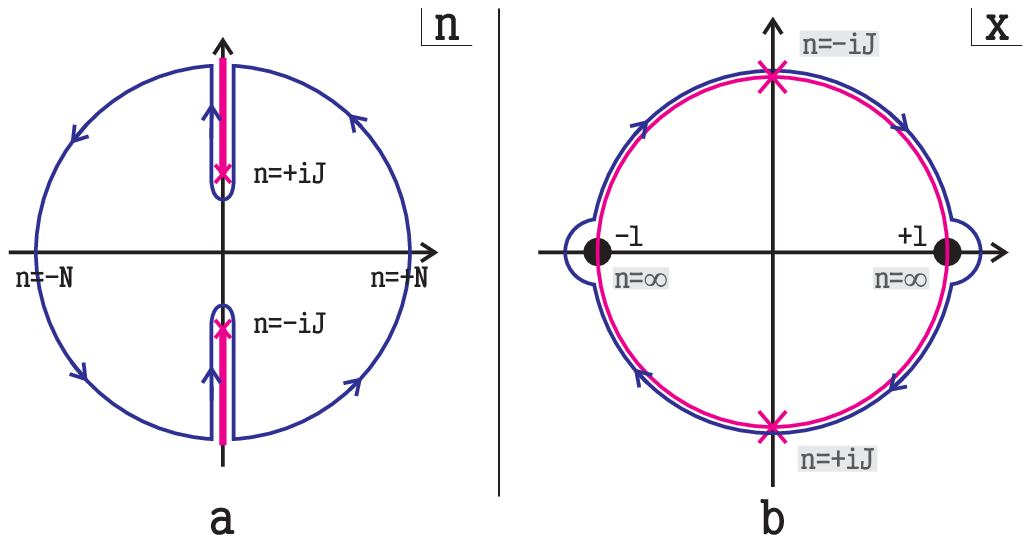}}
    \caption{\small\textsf{\textit{\textbf{\normalsize a.} Analytical structure of the BMN frequencies $\sqrt{n^2+\J^2}$ and integration contour for (\ref{cot}). \textbf{\normalsize b.}~Same picture in the $x$ plane obtained through the map $\frac{x}{x^2-1}=\frac{n}{2\J}$. The branchpoints are mapped to $x=\pm i$. As $N\rightarrow \infty$ the integration path is mapped to the unit circle.}}\label{fig:simple}
}
\end{figure}

Let us consider the sum of the BMN frequencies \cite{Berenstein:2003gb} $\sqrt{n^2+\J^2}$ from $n=-N$ to $n=N$ with $N$ large. In the $n$ plane, for each frequency, the integral (\ref{cot}) over $n$ can be deformed to run along the cuts in the imaginary $n$ axis with branchpoints $n=\pm i \J$ as depicted in figure 1a. Moreover, for large $\J$, we can replace the $\cot$ in (\ref{cot}) by $\pm i$ for the lower/upper half plane. Through  (\ref{map}) we can map the integration contour to the $x$ plane.
For this solution the quasi-momenta are very simple and
$$
p_i-p_j= \frac{4\pi \J x}{x^2-1}
$$
so that the branchcuts in the $n$ plane are mapped to the unit circle with the branchpoints $n=\pm i \J$ sent to $x=\mp i$ -- see figure 1b . The integration contour for large $N$ is mapped to the solid (blue) contour in figure 1b and tends to the unit circle as $N\to \infty$.

For general classical solutions the picture is similar. For large $n$ the fluctuation frequencies should behave like $\sqrt{n^2}$ so that we will always have two branchpoints like as figure $1a$. These branchpoints are always large if the classical solution has some large global charge \cite{GV2}. As before, when computing the integral along these branchcuts we can replace $\cot(\pi n)\to \mp i \,{\rm sign}({\rm Im}(n))$ with exponential precision. Let us call this contribution by $I_{phase}$ -- the solid (blue) contour in figure 2a.
Generically, contrary to what happened in the previous example, this is not the end of the story.
There could be additional cuts in the $n$ plane whose contribution we have to subtract in order to pick the poles in $\cot$ (and only these poles) in (\ref{cot}). We denote the contribution from these integrals over the new cuts in the $n$ plane by $I_{anomaly}$ -- the dashed (blue) contours in figure 2a. For the simple solutions considered in the literature these two terms usually bear the names of \textit{non-analytic} and \textit{analytic} contributions.
\begin{figure}[t]
    \centering
        \resizebox{151mm}{!}{\includegraphics{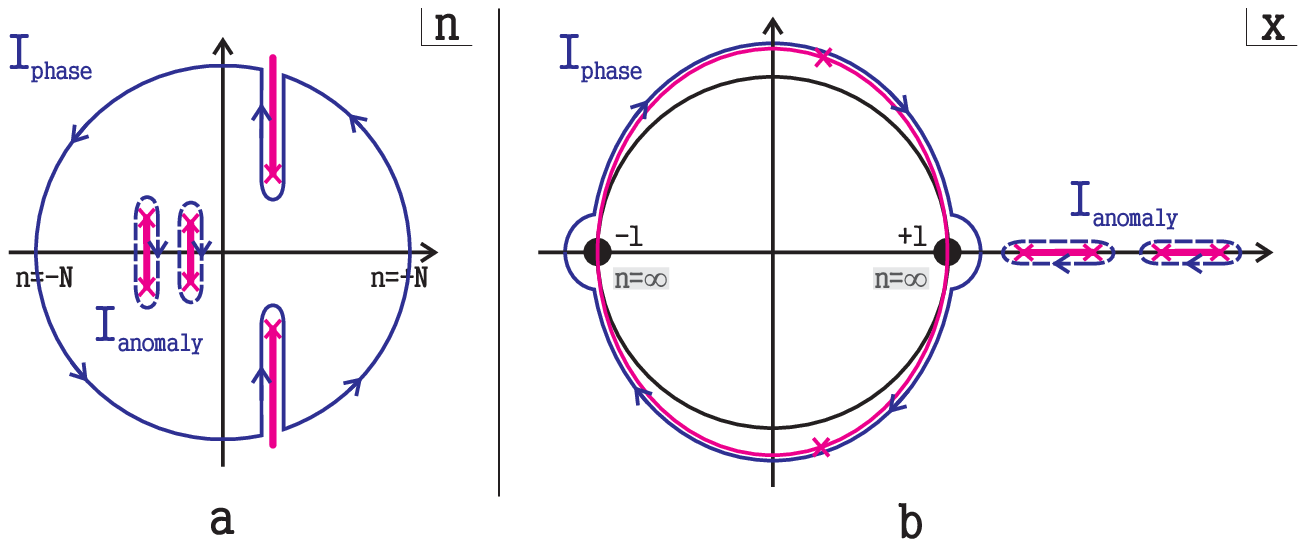}}
    \caption{\small\textsf{\textit{\textbf{\normalsize a.} Typical analytical structure of the excitation energies as a function of the mode number $n$. The branchpoints associated to the cuts going to infinity are large if some charge of the classical solution is large. There could also be extra cuts in the $n$ plane. The integral (\ref{cot}) can be then split into two contributions $I_{phase}$ and $I_{anomaly}$ as depicted in the figure. \textbf{\normalsize b.} The contour $I_{phase}$ going along the large cuts in the $n$ plane is mapped into some ellipsoidal form in the $x$ plane. The contours around the extra cuts in the $n$ plane are mapped to the cycles around the cuts of the classical solution around which we are quantizing.}}\label{fig:simple}
}
\end{figure}

The large branchcuts in the $n$ plane will then be mapped to some ellipsoidal curve in the $x$ plane passing through\footnote{For large $n$ the position of the pole should always be close to $x=\pm 1$ where all quasi-momenta have simple poles \cite{Arutyunov:2004yx}.} $x=\pm 1$ whereas the integrals around the extra cuts in the $n$ plane are mapped precisely to the integrals over the cuts of the original classical solution \cite{GV2} -- see figure 2b.
Finally, to pass from an integral over the $n$ plane to an integral in the $x$ plane we just need to use the map (\ref{map}),
\beq
dn=\frac{p_i'-p_j'}{2\pi}\,dx_n \,, \la{map2}
\eeq
so that we see the typical $p'$ and $\cot (\pi p)$ which always appear in the finite size correction analysis \cite{Schafer-Nameki:2005tn,Beisert:2005mqHernandez:2005nfBeisert:2005bvGromov:2005gp}! We will study the general finite size corrections in a forthcoming publication \cite{GV2} and elucidate its relation to $I_{anomaly}$.
In this paper we prove the universality of the Hernandez-Lopez (HL) scalar factor by analyzing the $I_{phase}$ contribution around \textit{any} classical configuration.

As we will see in the following section, the addition of the dressing factor $e^{i\cal V}$ in the middle equation \eq{middle} amounts to adding the potential to each quasimomenta $p_i\to p_i\pm {\cal V}/2$. Having this in mind, let us give a sketch of the proof. As we mentioned in the beginning of this section, by adding\footnote{And also image of this pole $
\frac{\alpha(1/x)}{1/x-x_n^{ij}}
$ according to $x\to 1/x$ symmetry \eq{xto1/x}.
}
$$
\frac{\alpha(x)}{x-x_n^{ij}}
$$
to the quasi-momenta $p_i$ and $p_j$ with $x_n^{ij}$ fixed by (\ref{map}) we are considering a quantum fluctuation with mode number $n$ and polarization $ij$. 
Then, if we want to get the contribution $I_{phase}$ we should integrate this pole over $n$ using (\ref{map2}) and sum over the different polarizations with the appropriate signs (\ref{cot}). For example for $\t p_1$ we have $(i,j)=(\t1,\t3),(\t1,\t 4)$ bosonic excitation coming with a plus sign and $(i,j)=(\t1,\h3),(\t 1,\h4)$ for the fermionic excitations summed with a minus sign -- see figure 3. Then we find that the contribution we must add for each $p_i$ is the same and reads
\beqa
{\cal V}(x)=\int_{-1}^{+1}\[\t p_3'+\t p_4'-\h p_3'-\h p_4'\]\(\frac{\alpha(x)}{x-y}-\frac{\alpha(1/x)}{1/x-y} \) \frac{dy}{2\pi}\la{calVxIntro} \,.
\eeqa
It is interesting to see that this potential is explicitly $x\to 1/x$ odd,
\beq
{\cal V}(x)=-{\cal V}(1/x)\,. \la{symV}
\eeq
Finally, as we will see in the next section, the combination of quasi-momenta appearing in \eq{calVxIntro} is precisely the one from which one reads the global charges $Q_n$,
\beq
\t p_3'+\t p_4'-\h p_3'-\h p_4'=\d_y\[G_4(y)-G_4(1/y)\] \,. \la{pG4}
\eeq
where
\beq
 G_4(y)=-\sum_{n=0}^\infty Q_{n+1} y^n \,, \la{GQ}
 \eeq
so that we can expand the denominators in (\ref{calVxIntro}) for large $x$ and obtain
 \beq
{\cal V}(x)=\alpha(x)\!\!\!\!\!
\sum_{\scriptsize\bea{c}r,s=2\\r\!+\!s\in Odd\eea}^\infty
\!\!\!\!
\frac{1}{\pi}\frac{(r-1)(s-1)}{(s-r)(r+s-2)}\(\frac{Q_r}{x^s}-\frac{Q_s}{x^r}\) \la{HL}
\eeq
where we recognize precisely the Hernandez-Lopez coefficients! To obtain the values of the potential for $|x|<1$ we can simply use the exact symmetry (\ref{symV}) which is not manifest in the form (\ref{HL}). In the next section we shall explain the tight relation between this potential $\mathcal{V}$ and the BS equations and provide a detailed derivation of the scalar factor.

\section{Constraining the scalar factor}
In this section we shall fill the gaps in the sketchy derivation above.
First we will start by explaining the relation between the quasi-momenta and the Beisert-Staudacher (BS) equations in the large $\lambda$ limit. We will see that relation (\ref{pG4}) follows immediately from the definition of the quasi-momenta and that the phase $\mathcal{V}(x)$ appearing in (\ref{middle}) is simply translated into a potential for the  quasi-momenta $p_i\to p_i\pm \mathcal{V}/2$. Then, we clarify the steps leading to (\ref{calVxIntro}), preform the large $N$ limit more carefully.

\subsection{Classical limit}
One of the main ingredients in the construction of the BS equations was the requirement that these equations reproduce the classical algebraical curve of \cite{BKSZ}.
For the sake of completeness, let us we make the passage to the classical limit explicit and at the same time introduce some important notations.

The seven BS equations, one for each node of a super Dynkin diagram of the $psu(2,2|4)$ algebra, give us the position of the roots $u_{a,j}$ where $a=1,\dots,7$ denotes the Dynkin node and $j=1,\dots,K_a$. The equation associated to the middle node of the Dynkin diagram takes the form (\ref{middle}).
In fact, for a given number of roots, the Bethe equations have several solutions. To classify them one takes the log of the Bethe equation associated with each root. The different choices of the branch of the log correspond to the different solutions. In other words, to each root $u_{a,j}$ one should associate a mode number $n_{a,j}$. Thus, a choice of mode numbers amounts  to fixing the quantum state.

To study the large $\lambda$ scaling limit,
\beq
u_j\sim  L\sim K_i\sim \sqrt{\lambda} \,,\la{LKlambda}
\eeq
of the BS equations it is useful to introduce $8$ functions $\{\h p_1,\h p_2,\h p_3,\h p_4,\t p_1,\t p_2,\t p_3,\t p_4\}$. First we define the resolvents $G_a$ and $H_a$ for each type of roots
\beqa
G_a(x)=\sum_{j=1}^{K_a}\frac{\alpha(y_{a,j})}{x-y_{a,j}} \,\, , \,\, H_a(x)=\sum_{j=1}^{K_a}\frac{\alpha(x)}{x-y_{a,j}} \la{GH} \,.
\eeqa
Then, denoting $\bar H_a(x)=H_a(1/x)$ and $\J=L/\sl$, we have
\beqa
\bea{l|l}
\!\h p_1\!=\!+ \displaystyle{\frac{2\pi \J x - G'_4(0)x }{x^2-1}}\!-\!H_1\!+\!H_2\!+\!\bar H_2\!-\!\bar H_3\!+\!\frac{1}{2}{\cal V}& \t p_1\!=\!+ \displaystyle{\frac{2\pi \J x  + G_4(0)}{x^2-1}}\!-\!H_1\!-\!\bar H_3\!+\!\bar H_4\!+\!\frac{1}{2}\cal V
{\color{white}\frac{\frac{1}{2}}{\frac{1}{2}}}\\
\!\h p_2\!=\!+\displaystyle{\frac{2\pi \J x - G'_4(0)x}{x^2-1}}\!-\!H_2\!+\!H_3\!+\!\bar H_1\!-\!\bar H_2\!+\!\frac{1}{2}\cal V &\t p_2\!=\!+\displaystyle{\frac{2\pi \J x  + G_4(0)}{x^2-1}}\!+\!H_3\!-\!H_4\!+\!\bar H_1\!+\!\frac{1}{2}\cal V
{\color{white}\frac{\frac{1}{2}}{\frac{1}{2}}}\\
\!\h p_3\!=\!-\displaystyle{\frac{2\pi \J x  - G'_4(0)x}{x^2-1}}\!-\!H_5\!+\!H_6\!+\!\bar H_6\!-\!\bar H_7\!-\!\frac{1}{2}\cal V&\t p_3\!=\!-\displaystyle{\frac{2\pi \J x  + G_4(0)}{x^2-1}}\!-\!H_5\!+\!H_4\!-\!\bar H_7\!-\!\frac{1}{2}\cal V
{\color{white}\frac{\frac{1}{2}}{\frac{1}{2}}}\\
\!\h p_4\!=\!-\displaystyle{\frac{2\pi \J x - G'_4(0)x}{x^2-1}}\!-\!H_6\!+\!H_7\!+\!\bar H_5\!-\!\bar H_6\!-\!\frac{1}{2}\cal V&\t p_4\!=\!-\displaystyle{\frac{2\pi \J x  + G_4(0)}{x^2-1}}\!+\!H_7\!+\! \bar H_5\!-\!\bar H_4\!-\!\frac{1}{2}\cal V
\eea \la{p}
\eeqa
\begin{figure}[t]
    \centering
        \resizebox{120mm}{!}{\includegraphics{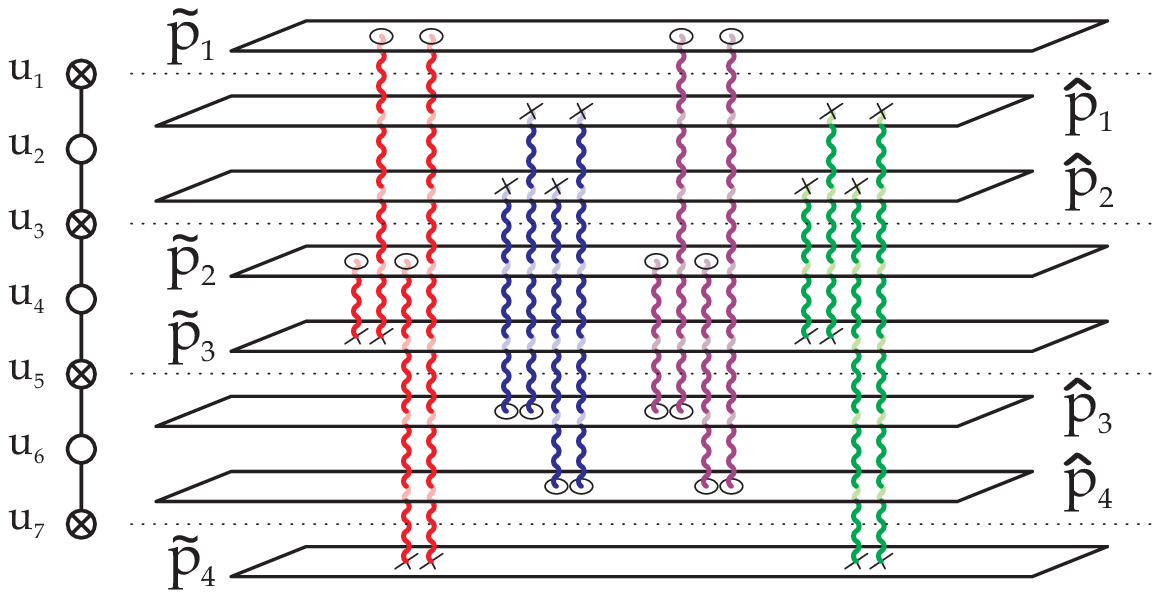}}
    \caption{\small\textsf{\textit{The 16 elementary physical excitations are the stacks (bound states) containing the middle node root. From the left to the right we have four $S^5$ fluctuations, four $AdS_5$ modes and eight fermionic excitations. The bosonic (fermionic) stacks contain an even (odd) number of fermionic roots signaled by a cross in the Dynkin diagram of $psu(2,2|4)$ in the left.}}\label{fig:simple}
}
\end{figure}
In the continuous limit, with a large number of roots for each mode number,  roots with the same $n_{a,j}$ will condense into square root cuts. Moreover roots belonging to consecutive nodes of the Dynkin diagram can form bound states and in this way give rise to a cut $\mathcal{C}_{ij}$ connecting non-consecutive Dynkin nodes.  As mentioned in the introduction only the middle roots $u_4$ carry energy (\ref{Delta4}). Then the 16 elementary physical excitations are the bound states represented in figure 1, each bound state corresponding to a different string polarization. These bound states, named stacks, were first found in \cite{BKSZ2} -- we refer to this article for more details. We denote the values of a function $p(x)$ above and below some of these cuts $\mathcal{C}_{ij}$ by $p^\pm$. Then, in the
large $\lambda$ limit, we can recast the seven BS equations as\footnote{In the notation of \cite{Beisert:2005fw} we use the grading $\eta_1=\eta_2=1$ corresponding to the Dynkin diagram of figure 3. Moreover we are considering the $1/\sl$ corrections coming from $\mathcal{V}\sim \mathcal{O}(1/\sqrt{\lambda})$ but we drop in this paper the finite size corrections usually called by anomaly terms. We shall discuss them separately \cite{GV2}.}
\begin{eqnarray}
p_i^+- p_j^-=2\pi n_{ij} \,\, , \,\, x\in \mathcal{C}_{ij} \la{pij2} \,.
\end{eqnarray}
From the definition of the quasi-momenta we can read the large and small $x$ asymptotics, the behavior at the $x=\pm 1 $ poles and understand how the several quasimomenta are related by the $x$ to $1/x$ inversion symmetry. To analyze the classical limit, moreover, we can drop the potential $\mathcal{V}$ whose contribution is of order $1/\sqrt{\lambda}$. We conclude that the analytical properties of $p$, together with the equations just mentioned are exactly the same as those of the quasimomenta defining the 8-sheet Riemann surface of Beisert, Kazakov, Sakai and Zarembo \cite{BKSZ}. Thus, the classical limit coincides with this continuous limit.

\subsection{Deriving the Hernandez-Lopez scalar factor}

Until the end of this section we drop the phase $\mathcal{V}$ in the BS equations (and thus also in (\ref{p})).
If we add a stack connecting sheets $i$ and $j$ to some configuration of Bethe roots with all roots condensed into some cuts as described above, the position of the new stack will be given by (\ref{map}) and
and all the other roots will be slightly shifted $u_j \rightarrow \t u_j$. Then the energy of the new configuration 
is given by the energy of the original configuration plus the fluctuation energy with mode number $n$ associated to the corresponding string polarization \cite{GV}
\beqa
\t\Delta=\Delta+\E_n^{ij} \,. \la{Delta3}
\eeqa
Let us now perform a simple rewriting exercise and treat each of the roots of this new stack separately in $p_k$. That is, if the stack contains a root associated with the Dynkin node $a$ we write
\beqa
G_a(x)\rightarrow G_a(x)+\frac{\alpha(x_n)}{x-x_n} \,\, , \,\, H_{a}(x) \to  H_{a}(x)+\frac{\alpha(x)}{x-x_n} \nn
\eeqa
where $G_a$ and $H_a$ are now defined with the sum over roots going only over $a=1,\dots,K_a$ where $K_a$ is the original number of roots of type $u_{a,j}$. Then, with this new stack, each quasi-momentum  $p_k$  can be written as before but using the new resolvents $G_a$ and $H_a$ containing only the $K_a$ original roots plus an extra term $V_k^{ij}$ which we call potential and read
\footnote{For example, consider a fermionic stack $i,j=\t2, \h3$ connecting $\tilde p_2$ and $\h p_3$.
As we see from figure 3 this stack is made of two almost coincident $u_4$ and $u_5$ roots. The first term in the potentials comes from the resolvent of the middle node though the $G_4(0)$ and $G_4'(0)$ terms present in all quasimomenta (\ref{p}). The new terms in $\t p_1,\t p_2,\h p_3, \h p_4$ come from the resolvents $H_4$ and $H_5$ which, for the other quasimomenta, are either not present or appear with opposite signs.}
\beq
\(
\bea{c}
V_{\h 1}(x)\\
V_{\h 2}(x)\\
V_{\h 3}(x)\\
V_{\h 4}(x)
\eea\)^{ij} =\(
\bea{c}
+1\\
+1\\
-1\\
-1
\eea\)\frac{x}{x^2-1}\frac{\alpha(x_n^{ij})}{(x_n^{ij})^2}
+\(
\bea{c}
+\delta_{\h1i}\\
+\delta_{\h2i}\\
-\delta_{\h3j}\\
-\delta_{\h4j}\\
\eea
\)\frac{\alpha(x)}{x-x_n^{ij}}
-
\(
\bea{c}
+\delta_{\h2i}\\
+\delta_{\h1i}\\
-\delta_{\h4j}\\
-\delta_{\h3j}\\
\eea
\)\frac{\alpha(1/x)}{1/x-x_n^{ij}}\la{Vh} \,,
\eeq
and
\beq\la{Vt}
\(
\bea{c}
V_{\t 1}(x)\\
V_{\t 2}(x)\\
V_{\t 3}(x)\\
V_{\t 4}(x)
\eea\)^{ij} =
\(
\bea{c}
-1\\
-1\\
+1\\
+1
\eea\)\frac{1}{x^2-1}\frac{\alpha(x_n^{ij})}{x^{ij}_n}-
\(
\bea{c}
+\delta_{\t1i}\\
+\delta_{\t2i}\\
-\delta_{\t3j}\\
-\delta_{\t4j}\\
\eea
\)\frac{\alpha(x)}{x-x_n^{ij}}
+
\(
\bea{c}
+\delta_{\t2i}\\
+\delta_{\t1i}\\
-\delta_{\t4j}\\
-\delta_{\t3j}\\
\eea
\)\frac{\alpha(1/x)}{1/x-x_n^{ij}} \,,
\eeq
In (\ref{p}) we saw that the inclusion of the phase $\mathcal{V}$ in the middle node of the BS equations amounts to
adding the same potential $\cal V$ to all quasi-momenta.
The main difference to what we have there is that now the $x_n$ roots hidden in the potential also contribute to the charges (because every stack contains a $u_{4}$ root)
\beq
Q_m=\oint \frac{dx}{2\pi i}\,\frac{G_4(x)}{x^m}+ \frac{\alpha(x_n)}{x_n^m} \,. \la{charges}
\eeq
Moreover the potentials $V_k^{ij}$ are different for different quasi-momenta.

Suppose that instead of (\ref{Delta3}) we want
\beq
\t \Delta=\Delta + I_{phase} \la{Delta5}
\eeq
with\footnote{As it is discussed in \cite{GV} in order to define this quantity unambiguously a precise prescription for the labeling of the fluctuation energies is needed. This point is discussed in Appendix A.}
\beq
I_{phase}\equiv\frac{1}{2}\lim_{N\to \infty}\int\limits_{-N}^N \sum_{ij} (-1)^F \E^{ij}_n dn \la{Eint}
\eeq
where the contour in the $n$ plane is as depicted in figure 4a. Then, by linearity, we need only to replace $V_k^{ij}$ by
\beq
V_k(x)=\frac{1}{2}\lim_{N\to \infty}\sum_{ij}\int\limits_{-N}^N  (-1)^F V_k^{ij}(x,x^{ij}_n) dn \la{Vs}\ .
\eeq
Let us now show that all the potentials $V_k$ are the same up to a sign and are equal to $\cal V$ \eq{calVxIntro} from the previous section. Indeed
\begin{enumerate}
\item{For each summand in \eq{Vs} we can pass to the $x$ plane through (\ref{map2}). As we explain in Appendix A we can assume that for all fluctuation energies the corresponding integral in the $x$ plane is the same and goes over the upper and lower halves of the circle of radius $1+\epsilon$ centered at the origin as plotted in figure 4b.}
\begin{figure}[t]
    \centering
        \resizebox{120mm}{!}{\includegraphics{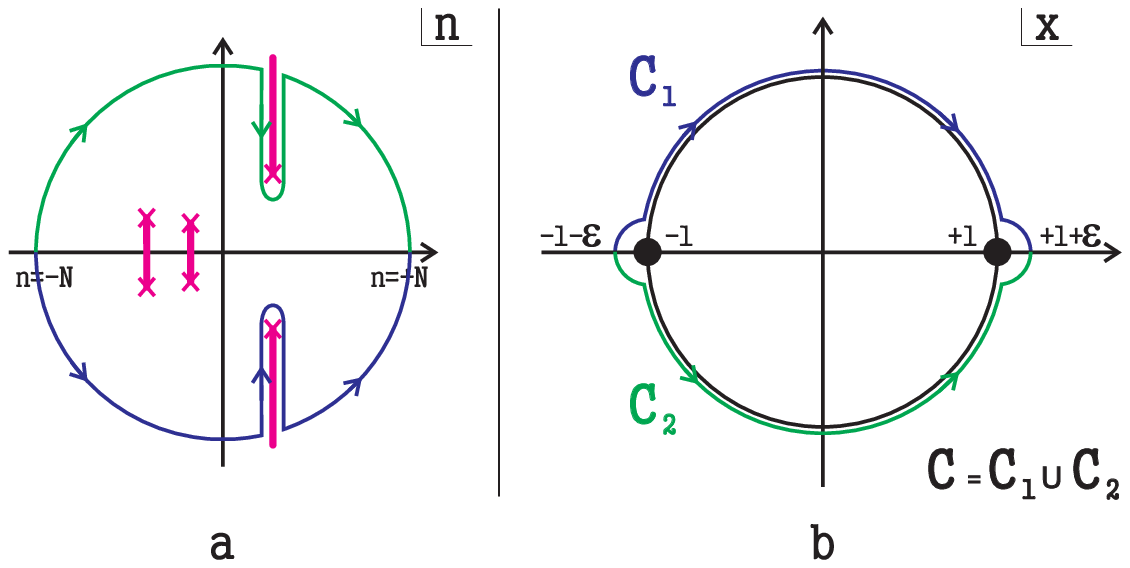}}
    \caption{\small\textsf{\textit{\textbf{\normalsize a.} The ``non-analytic" contribution $I_{phase}$ is given by the integral (\ref{Eint}) whose integration path goes along the large cuts discussed in section 2. The difference in orientations with respect to figure 2a is due to the absence of $\cot$ in expression (\ref{Eint}) compared to (\ref{cot}). \textbf{\normalsize b.} In the $x$ plane the integral can safely be deformed to go over the upper and lower halves of the unit circle. In the main text we use the shorthand $\int_{-1}^{+1}$ to denote $\frac{1}{2}\int_{C_1}+\frac{1}{2}\int_{C_2}$. The relation between the large $N$ regularization in the $n$ plane and the $\epsilon$ regularization in the $x$ plane is discussed in appendix A.}}\label{fig:simple2}
}
\end{figure}
\item{The first terms in (\ref{Vh}) and (\ref{Vt}) do not contribute to $V_k$. Indeed, if we integrate some function of $x_n^{ij}$ summed over the $16$ possible excitations listed in figure 3 with a $(-1)^F$ weight
\beq
\sum_{ij} (-1)^F \int_{-N}^{N}  f(x_n^{ij}) dn  \nn
\eeq
we obtain, using (\ref{map2})\footnote{We can as well use the quasi-momenta with the resolvents $G_a$ and $H_a$ summed only over the original roots because the inclusion of the potentials in (\ref{map2}) is an higher order effect.},
\beq
\int_{-1-\epsilon}^{+1+\epsilon} f(y) \[\sum_{i=1,2\,, j=3,4} (\t p'_i-\t p'_j)+(\h p'_i-\h p'_j)-(\t p'_i-\h p'_j)-(\h p'_i-\t p'_j)\] \frac{dy}{2\pi}=0\,.
 \la{toint2}
\eeq }
\item{Finally, consider for example $V_{\h 1}$. We have
\beqa
V_{\h 1}(x)&=&\frac{1}{2}\int_{-1-\epsilon}^{1+\epsilon}\[(\h p_1'-\h p_3')+(\h p_1'-\h p_4')-(\h p_1'-\t p_3')-(\h p_1'-\t p_4')\]\frac{\alpha(x)}{x-y} \frac{dy}{2\pi} \nn\\
&-&\frac{1}{2}\int_{-1-\epsilon}^{1+\epsilon}\[(\h p_2'-\h p_3')+(\h p_2'-\h p_4')-(\h p_2'-\t p_3')-(\h p_2'-\t p_4')\]\frac{\alpha(1/x)}{1/x-y} \frac{dy}{2\pi}\nn\ .
\eeqa
We see that $\h p_1$ and $\h p_2$ drop out so that the expression simplifies considerably. The same happens
for the other $V_{i}$ and moreover, due to the super-tracelessness of the monodromy matrix, $\t p_1+\t p_1+\t p_1+\t p_4=\h p_1+\h p_1+\h p_1+\h p_4$, and all the potentials are equal. Using (\ref{pG4}) we have
\beqa
V_{\h 1,\h 2,\t1,\t2}(x)=-V_{\h 3,\h 4,\t3,\t4}(x)\equiv{\cal V}(x)=\int_{-1}^{1}\partial_{y}\[G_4(y)-G_4(1/y)\]\(\frac{\alpha(x)}{x-y}-\frac{\alpha(1/x)}{1/x-y} \) \frac{dy}{2\pi}\la{calVx} \,.
\eeqa
}
\end{enumerate}
Notice also that due to 3) the extra terms in the charges (\ref{charges}) give no contribution! Thus, since now all potentials $V_k$ are equal to $\cal V$, we proved that for arbitrary configuration of Bethe roots, the addition of the Hernandez-Lopez phase will lead to an energy shift given by $I_{phase}$.

As we saw in the previous section, to obtain the Hernandez-Lopez phase as usually written in terms of charges it suffices to use (\ref{GQ}). If we want, on the other hand, to write $$e^{i\mathcal{V}(y_{4,k})}=\prod\limits_{j\neq k}^{K_4} e^{i\theta(y_{4,k},y_{4,j})}$$ where the factorized scattering property is manifest we just need to use the definition (\ref{GH}) and integrate over $y$ to get\footnote{By resuming the Hernandez-Lopez coefficients the phase $\theta(x,y)$ was written down in \cite{Arutyunov:2006iu}, see also the appendix B in \cite{Schafer-Nameki:2006ey}.}
\beqa
\theta(x,y)=-\frac{\alpha(x)\,\alpha(y)}{\pi}\[\(\frac{1}{(x-y)^2}+\frac{1}{(xy-1)^2}\)\log\(\frac{x+1}{x-1}\frac{y-1}{y+1}\)+\frac{2}{(x-y)(xy-1)}\] \la{theta}
\eeqa
The \textit{real} scattering phase, the phase that describes the scattering between two magnons in the Bethe ansatz equation, must inherit the explicit $x$ to $1/x$ oddness (\ref{symV}) of the potential. To obtain the values of the phase for $|x|<1$ we use $\theta(1/x,y)=-\theta(x,y)$. Alternatively, we recall that the contour in figure $4b$ tells us that to be completely rigorous we should replace the $\log$ in (\ref{theta}) by $\frac{1}{2}\(\log_{+}(\dots)+\log_{-}(\dots)\)$ where $\log_{\pm}$ has a branchcut in the upper/lower half of the unit circle -- see figure 4b. Then the expression for $\theta(x,y)$ becomes explicitly $x$ to $1/x$ odd and is discontinuous on the unit circle.
If, on the other hand, we analytically continue the expression (\ref{theta}) from some point $x$ outside the unit circle up to some point $1/x$ inside the unit circle we get $2\pi i$ from one of the $\log_{\pm}$ so that we trivially find
\beqa
i\theta(x,y)+i\t\theta(1/x,y)=-\alpha(x)\,\alpha(y)\(\frac{1}{(x-y)^2}+\frac{1}{(xy-1)^2}\)\,, \nn
\eeqa
which is precisely
Janik's crossing relation \cite{Janik:2006dc} for the dressing factor at $1/\sl$ order \cite{Arutyunov:2006iu}.

\section{Speculations on nesting}
There are several indications pointing towards the existence of an extra hidden level in the Bethe ansatz equations. In \cite{Mann:2005ab,GKSV} the classical algebraic curve for the string moving in $S^5\subset AdS_5\times S^5$ was obtained as the classical limit of the quantum nested Bethe ansatz equations coming from the Zamolodchikov's bootstrap procedure \cite{Zamolodchikov:1977nu} where in addition to the magnons described by the roots $u_j$ we have the rapidities $\theta_\alpha$ of the relativistic particles with $O(N)$ isotopic degree of freedom. In \cite{Rej:2005qt} it was observed that the BDS equations \cite{Beisert:2004hm} mentioned in the introduction (\ref{su2Gauge}) could be obtained from the Hubbard model where the electron has a spin which can create spin waves described by the roots $u_j$, but also has momentum $p_\alpha$. In both cases the introduction of the extra level simplifies the structure of the Bethe equations considerably.
Moreover, in the same setup as in \cite{GKSV} it was also found \cite{Gromov:2006cq} that the elimination of the rapidities $\theta_\alpha$ from the simple Bethe equations would lead to the more complicated AFS equations (\ref{su2String}).
More recently it was argued \cite{Rej:2007vm} that the conjectured all loop dressing factor should also bear its origin from an extra level in the Bethe ansatz equations\footnote{In \cite{Sakai:2007rk} it was argued that the dressing factor could instead come from a dressing of the bare vacuum in the original \cite{Beisert:2005fw} equations with no scalar factor at all.}.
\begin{figure}[t]
    \centering
        \resizebox{120mm}{!}{\includegraphics{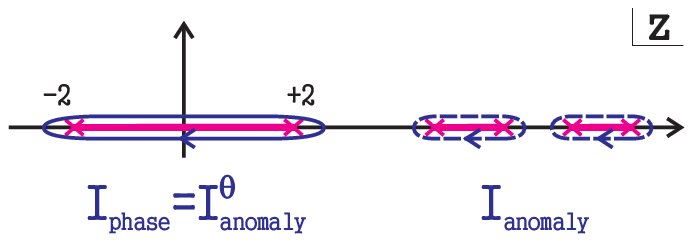}}
    \caption{\small\textsf{
{On $z$ plane the contributions $I_{phase}$ and $I_{anomaly}$ can be treated in absolutely equal footing. This hints the existence of an extra level in the Bethe ansatz equations whose finite size corrections would give the contribution $I_{phase}$. }}}\label{fig:simple2}
\end{figure}
Thus we would like to understand if (\ref{calVx}) could have imprinted signs of a nested Bethe ansatz structure. Since the generalization of the Bootstrap method for the supercoset $PSU(2,2|4)$ is not known we will proceed at a rather speculative and schematic level.
Let us recall that in the scaling limit the $S^5$ nested Bethe ansatz equations read (see e.g. equations 6.3 or 3.15 in \cite{GKSV})
\beqa
&&\sG_\theta+K_{\theta a} G_a=2\pi m \,\,\ \ \  \,\,\, \ \ \ \quad, \,\, z\in [-2,2] \la{lulu} \\
&&\sG_a+K_{a b} G_b +K_{a \theta} G_\theta=2\pi n_a \,\, , \,\, z\in \mathcal{C}_a \nn
\eeqa
where the $K$'s are numeric coefficients and where the cut from $[-2,2]$ is the image of the unit circle under the Zhukovsky map. The discontinuity of $G_\theta$ is related to the density of the extra level particles $\theta_\alpha$ and is given on the $x$ plane by, roughly speaking,
\beq
\rho_\theta=-\frac{p_2(x)-p_2(1/x)}{2\pi i}\ . \la{prho}
\eeq
Suppose we want to take into account $1/\sl$ corrections coming from the extra level. We denote the extra term in the first equation in (\ref{lulu}) by $\frac{1}{\sl} V(\theta,\{u\})$. Then, if we want to eliminate $\theta$'s from this equation and plug them into the second one to obtain some effective equation for the magnons as in \cite{Gromov:2006cq}, we must solve $\delta\, \sG_\theta=\frac{1}{\sl} V(\theta)$ and plug the solution of this Hilbert problem into the second equation. In the second equation this will then appear as an extra phase. The Hernandez-Lopez phase seems to fit into this construction.
Indeed, if we go to the Zhukovsky plane in (\ref{calVx}) through $z=x+1/x$ and $w=y+1/y$ we will have
\beqa
{\cal V}(x)=\frac{4\pi}{\sqrt{\lambda}}\int_{-2}^{2} V(w)\frac{\sqrt{4-w^2}}{\sqrt{z^2-4}}\frac{dw}{w-z}  \,\, , \,\, V(w)=\frac{\partial_{w}\[G_4(y)-G_4(1/y)\]}{2\pi i} \nn
\eeqa
which precisely indicates that it came from the solution of a Hilbert problem with $V(w)$ resembling the derivative of the density in (\ref{prho}). Indeed
\beq
V(w)=\partial_{w}\[-\frac{\t p_2(y)-\t p_2(1/y)}{2\pi i}+\frac{\h p_2(y)-\h p_2(1/y)}{2\pi i}\] \nn
\eeq

As explained in section 2 we are considering classical solutions with some large conserved charge.
Then the quasi-momenta $p(x)$ scales like that charge close to the unit circle \cite{GV2} and so will the density (\ref{prho}). Thus if we assume that the anomaly for these roots is of the usual form $\rho_\theta'\coth \pi \rho_\theta$ we see that we can drop the cotangent with exponential precision and we are left precisely with the derivative of the density of $\theta$ particles which, as we argued above, strongly resembles $V(w)$! Moreover, the integral over the unit circle in 2b will be mapped to the cycle around the $\theta$ from $z=-2$ to $z=2$ thus leading to the very democratic figure 5.

\section{Conclusions}
The fluctuation energies around any classical solution can be computed from the classical algebraic curve \cite{GV} which, a priori, contains no information about the HL phase suppressed as $1/\sl$.
On the other hand, if we expand the energy of some state in $1/\sl$ we will obtain the classical energy of order $\sqrt{\lambda}$ plus a finite correction which is known \cite{Frolov:2002av} to be the sum of these fluctuation frequencies. In other words, the sub-leading order is constrained by the leading order!

This interconnection between the leading and sub-leading terms means that, if one believes in the existence of a Bethe ansatz description of the quantum string, then the first correction to the dressing factor is completely constrained.
Thus we might wonder if this procedure can be iterated to fix order by order, the full scalar factor.

In this paper we found that the first correction to the dressing factor must have the form
\beqa
{\cal V}(x)=\int_{-1}^{1}\partial_{y}\[G_4(y)-G_4(1/y)\]\(\frac{\alpha(x)}{x-y}-\frac{\alpha(1/x)}{1/x-y} \) \frac{dy}{2\pi}\nn \,,
\eeqa
in order to accommodate for the $1/\sl$ ``non-analytical" part of the $1$-loop shift for \textit{any} classical configuration.

One can see that the $1/\sl$ corrections in the Beisert-Staudacher equations with the Hernandez-Lopez phase to any conserved charge $Q_n$ is given by one half of the graded sum of the corresponding charges of the fluctuations -- as it was the case for energy.

Moreover, this representation of the phase has several interesting features. First of all, it is extremely simple -- the Hernandez-Lopez coefficients follow after a trivial expansion of the resolvent in conserved charges and to find the scattering phase $\theta(x,y)$ between two magnons one merely needs to plug the definition of the resolvent into the integral without the need to perform any re-summation. By construction, the cut structure is also very clear and thus the crossing relation becomes transparent.

Finally this phase has imprinted signs pointing towards the existence of an extra level of $\theta$ roots (which are in fact rapidities of physical particles of the theory) in the Bethe ansatz equations. These extra roots live in the unit circle and in the quasi-classical limit condense into some smooth distribution whose density resembles the $G_4(y)-G_4(1/y)$ \cite{GKSV} appearing in $\mathcal{V}(x)$. It is very likely that the anomaly associated with the roots of this new level naturally reproduce the Hernandez-Lopez phase.

\subsection*{Acknowledgements}
We would like to thank K.~Zarembo and V.~Kazakov for many useful discussions.  The work of
N.G. was partially supported by French Government PhD fellowship,
by RSGSS-1124.2003.2 and by RFFI project grant 06-02-16786. P.~V.
is funded by the Funda\c{c}\~ao para a Ci\^encia e Tecnologia fellowship
{SFRH/BD/17959/2004/0WA9}.

\section*{Appendix A -- Large $N$ limit}

In the $x$ plane the contour in figure 4a is mapped to that in figure 4b. For large $N$ the contour starts at $-1-\epsilon_-^{ij}(N)$ and ends at $+1+\epsilon_+^{ij}(N)$. In this appendix we perform a careful analysis of the large $N$ limit.
\subsection*{A.1 Asymptotics of quasimomenta and expansion of $x_n$}
Large $n$'s are mapped to the vicinity of $\pm 1$ where
\beqa
\h p_{2}\simeq+\frac{\alpha_\pm}{x\mp 1}+\sum_{n=0}\h a_n^\pm(x\mp 1)^n\nn\,\, , \,\,
\t p_{2}\simeq+\frac{\alpha_\pm}{x\mp1}+\sum_{n=0}\t a_n^\pm(x\mp1)^n \,, \\
\h p_{3}\simeq-\frac{\alpha_\pm}{x\mp1}+\sum_{n=0}\h b_n^\pm(x\mp1)^n\,\, , \,\,
\t p_{3}\simeq-\frac{\alpha_\pm}{x\mp1}+\sum_{n=0}\t b_n^\pm(x\mp1)^n \,. \nn
\eeqa
The remaining quasimomenta are fixed by the $x\rightarrow 1/x$
symmetry
\beqa\nn
\t p_{1,2}(x)&=&-2\pi m-\t p_{2,1}(1/x)\\ 
\t p_{3,4}(x)&=&+2\pi m-\t p_{4,3}(1/x)\la{xto1/x}\\
\h p_{1,2,3,4}(x)&=&-\h p_{2,1,4,3}(1/x)\,.\nn
\eeqa

From this expansion we can read the large $n$ behavior of $x_n^{ij}$ defined by (\ref{map}). Let us, however,
use a more general definition
\beq
p_i(x_n^{ij})-p_j(x_n^{ij})=2\pi (n-m_i+m_j).
\eeq
For $n\rightarrow \pm\infty$ all $x_n^{ij}$ are close to $\pm 1$ and we find
\beq
x_n^{ij}=\pm 1+\frac{\alpha^{\pm}}{\pi n}+\O\(1/n^2\)
\eeq
where we notice that the first $1/n$ coefficient is universal and fixed uniquely by the residues of the quasi-momenta.

\subsection*{A.2 Large $N$ versus $\epsilon$ regularization}
The main goal of this appendix is to justify the integration path used in the main text where for all $ij$ the
integral in the $x$ plane
starts from $-1-\epsilon$ and ends at $1+\epsilon$ as depicted on the figure \ref{fig:simple2}b.
However by definition \eq{Vs} we have to start from the large $N$ regularization.
These two regularization in principal are not equivalent, since $x_N^{ij}$'s
are not exactly equal for all $ij$ and thus we should calculate the difference between both regularizations. For example
\beq
\delta V^{reg}_k(x)=\frac{1}{2}\lim_{N\to \infty,\epsilon\to 0}
\sum_{ij}\(\int\limits_{x^{ij}_{-N}}^{x^{ij}_{N}}-\int\limits_{-1-\epsilon}^{1+\epsilon}\)  (-1)^F V_k^{ij}(x,y)\frac{p'_i(y)-p'_j(y)}{2\pi} dy \la{VsA} \,.
\eeq
The difference of the integrals above can be rewritten as a sum of two integrals, one from $x_{-N}^{ij}$ to $-1-\epsilon$ and another from $1+\epsilon$ to $x_{N}^{ij}$, and thus we can use expansion of quasimomenta
around $\pm 1$ to evaluate this $\delta V^{reg}_k(x)$. One can see that only first terms in (\ref{Vh},\ref{Vt}) could be responsible for a non-zero difference given by
\beqa\nn
\delta V^{reg}_{\h 1,\h 2}(x)=-\frac{\pi^2 x}{(x^2-1)\sqrt\lambda}\(\frac{1}{\alpha_+}+\frac{1}{\alpha_-}\)\(m+m_{\h 1}+m_{\h 2}-m_{\t 1}-m_{\t 2}\)\(m+m_{\t 3}+m_{\t 4}-m_{\h 3}-m_{\h 4}\)\\
\delta V^{reg}_{\t 1,\t 2}(x)=-\frac{\pi^2 x}{(x^2-1)\sqrt\lambda}\(\frac{1}{\alpha_+}-\frac{1}{\alpha_-}\)\(m+m_{\h 1}+m_{\h 2}-m_{\t 1}-m_{\t 2}\)\(m+m_{\t 3}+m_{\t 4}-m_{\h 3}-m_{\h 4}\) \,. \nn
\eeqa
If the potentials $V_k$ are to be originated from a phase in the middle note of the BS equations then they should all be equal. This means that in order to be consistent with the phase origin, these two terms should be zero. Fortunately
it is possible to choose $m_i$ in such a way that it is so. For example
\beq
m_{\t1}=m\,\,,\,\,m_{\t4}=-m
\eeq
and all the others $m_i$ are zero. This amounts to a different prescription  for the mode numbers comparatively to \cite{GV}. For obvious reasons let us denote it by Bethe ansatz friendly prescription. Contrary to what we had in \cite{GV} we have no obvious argument, except the obvious Bethe ansatz friendliness, in favor of this new prescription.
For the $sl(2)$ and $su(2)$ one cut solutions this prescription gives the same result (with exponential precision in large $\cal J$) as in \cite{Frolov:2003qcFrolov:2003tuArutyunov:2003zaPark:2005jiBeisert:2005cw,Schafer-Nameki:2005tn,Sakura}.

By the same means we can see that in \eq{charges} the last term does not contribute in the Bethe ansatz friendly prescription.

\end{document}